\documentstyle[12pt,aaspp4]{article}
\begin{document}

\title{Spatially Resolved Observations of the Galactic Center Source, IRS 21}

\author{A. Tanner, A. M. Ghez\altaffilmark{1}, M. Morris and E. E. Becklin}
\affil{UCLA Department of Physics and Astronomy, Los Angeles, CA 90095-1562}
\author{A. Cotera} 
\affil{Steward Observatory, University of Arizona, Tucson, AZ 85721}
\author{M. Ressler, M. Werner} 
\affil{Jet Propulsion Lab, 169-327, 4800 Oak Grove Drive, Pasadena, CA, 91109} 
\author{P. Wizinowich}
\affil{W.M. Keck Observatory, 65-1120 Mamalahoa Hwy., Kamuela, HI 96743} 
\authoremail{tanner@astro.ucla.edu}

\begin{abstract}

We present diffraction-limited 2-25 $\micron$ images obtained with the W. M. Keck 10-m telescopes 
that spatially resolve the cool source, IRS 21, one of a small group of enigmatic objects in the central parsec of our Galaxy 
that have eluded classification. 
Modeled as a Gaussian, the azimuthally-averaged intensity profile of IRS 21 has 
a half-width half-maximum (HWHM) size of 650$\pm$80 AU at 2.2 $\micron$ and an average HWHM size 
of 1600$\pm$200 AU at mid-infrared wavelengths. These large apparent sizes imply an extended 
distribution of dust. The mid-infrared color map indicates that IRS 21 is a 
self-luminous source rather than an externally heated dust clump as originally suggested. 
The spectral energy distribution has distinct near- and mid-infrared components. 
A simple radiative transfer code, which simultaneously fits the near- and mid- infrared photometry and intensity profiles, 
supports a model in which the near-infrared radiation is scattered and extincted light from an embedded central source, while 
the mid-infrared emission is from thermally re-radiating silicate dust. 
We argue that IRS 21 (and by analogy the other luminous sources along the Northern Arm) is a 
massive star experiencing rapid mass loss and plowing through the Northern Arm, thereby generating a bow shock, which is spatially
resolved in our observations.

\end{abstract}

\altaffiltext{1}{Also affiliated with UCLA Institute for Geophysics and Planetary Physics}

\section{Introduction}

IRS 21 is one of the most distinctive objects within the central parsec of our Galaxy, a region abounding in extraordinary objects. 
Within the central parsec, IRS 21 has the largest K-band [2.2 $\micron$] polarization, $\sim$10\% (Ott et al. 1999; Eckart et al.
1995), and is one of the few objects to have both extremely red colors (H-K$>$3)
and a featureless K-band spectrum (Blum, Sellgren, \& Depoy 1996; Krabbe et al. 1995; Ott et al. 1999). 
Gezari et al. (1985) initially identified IRS 21 in 8.3 and 12.4 $\micron$ 
images as a strong mid-infrared peak located along the Northern Arm, a tidal 
stream of dust and gas (Gezari et al. 1985, Lacy et al. 1980) that is infalling towards and orbiting around the 
supermassive (M$_{BH}$ $\sim$ 2.6$\times$10$^6$ M$_{\odot}$) 
central black hole Sgr A$^*$ (Genzel et al. 1996, 2000; Ghez et al. 1998, 2000). 
Along the Northern Arm, there are a handful of other similarly strong mid-infrared sources (L$\sim$10$^{4-5}$L$_{\sun}$), 
including IRS 1, 2, 5, 8, and 10 (nomenclature from Becklin et al. 1978), which will be referred to collectively as the 
``Northern Arm Sources.''  In addition to having similar mid-infrared luminosities, these objects share the characteristics 
of featureless near-infrared spectra\footnote{Only three of the Northern Arm sources, IRS 21, 
1W and 10W, have measured K-band spectra (Krabbe et al. 1995).} and relatively cool colors. 
Although there are only a few other comparably luminous
objects in the central parsec, IRS 21 and the other Northern Arm sources
have eluded definitive classification. 

Initially, Gezari et al. (1985) suggested that IRS 21 is an externally
heated, high-density dust clump. Several other classifications have 
also been proposed since then, including an embedded early-type star and a 
protostar (Krabbe et al. 1995; Blum et al. 1996; Ott et al. 1999). 
The most recent analysis of this object was presented by Ott et al., 
who marginally resolved IRS 21 in 2.2 $\micron$ images with an angular resolution of 0$\farcs$15.
Although they do not report a physical size, they use 
the featureless, red spectra, along with the relative orientations
of the 2 and 12 $\micron$ polarization angles, to argue that IRS 21 is a 
recently formed star still embedded in its protostellar dust
shell.

The suggestion that IRS 21 is a newly formed object would favor a scenario wherein all the Northern 
Arm sources originated within the Northern Arm. At present, however, establishing a dynamical 
association between the gas and the infrared sources is not possible, as the V$_{LSR}$ for these sources has not been determined explicitly; their 
near-infrared spectra are featureless, and the proper motion measurements of the stars and gas at their position are insufficient.
There is, however, a serious problem with the protostellar hypothesis even without the velocity data;
the dynamical timescale for the gas along the Northern Arm is as short as 10$^4$ years (Lo and Claussen 1983),
substantially shorter than the timescale for star formation. Therefore, the Northern Arm sources, if they
are young, would have to have formed, or to have begun forming, prior to the infall of the Northern Arm.
In this case one would not expect the stars to still be embedded in the gas of the 
Northern Arm given that  
the newborn stars would have immediately started following ballistic orbits while the gas in which 
they formed is subject to strong, additional, non-gravitational forces.  Unlike the stars, the gas is subject to forces such as: 1) magnetic forces
in the strongly magnetized medium (Aitken et al. 1998), 2) the ram pressure of the strong
stellar winds emanating from the central cluster of luminous emission-line stars (Krabbe et al. 1995; Paumard et al. 2001), 
3) the radiation pressure from these same stars, 4) the gas dynamical pressures arising from the Northern Arm
being ionized on one side by radiation from the central star cluster, and 5) the forces accompanying 
shocks from the collisions of gas streams such as the Northern Arm and the Circumnuclear Disk. The  divergence of forces experienced by the stars 
and gas is likely most pronounced for IRS 21, which is at the apparent leading edge of this train of sources. 

An alternative hypothesis to having the Northern Arm sources form within the 
Northern Arm, while still allowing them to be young, is to admit that they are 
members of the central parsec cluster of apparently young, luminous 
emission-line stars, but that they be fortuitously embedded
within or superimposed upon the Northern Arm.
Based on $\sim$10\% coverage of the central parsec by the Northern Arm, however,
the chance\footnote{ Within the central parsec, six of the eight
strong mid-infrared sources lie along the Northern Arm.} that these are intrinsically mid-infrared sources that are
accidently superimposed on the Northern Arm is less than 10$^{-4}$.

These considerations lead us to suppose that the Northern Arm sources may be
ambient, luminous stars of the Galactic Center population that are presently
embedded in the Northern Arm. In order to assess this hypothesis, we have undertaken a study of the 
photometry and spatially resolved structure of IRS 21 at near- and mid-infrared wavelengths,
using diffraction limited imaging with the W.M. Keck 10-m telescopes. 
At the distance to the Galactic Center, 8 kpc (Reid 1993), 
the angular resolution of these images at the wavelength extremes of this study 
(0$\farcs$05 at 2.2 $\micron$ and 0$\farcs$31 at 24.5 $\micron$) 
correspond to 400 AU and 4960 AU, respectively, and are a factor of 2-3 
times better than previous measurements (Ott et al. 1999; Gezari 1992).

\section{Observations \& Data Reduction}

\subsection{Near-Infrared Images} 

Speckle imaging observations of IRS 21 were obtained in the K [2.2 $\micron$] bandpass using the W. M. Keck I 10-meter telescope and the facility 
near-infrared camera (NIRC; Matthews \& Soifer 1994; Matthews et al. 1996)
on the nights of 1995 June 10-12, 1996 June 26-27, 1997 May 13, 1998 April 2, 1998 May 14-15, 1998 August 4-6, 1999 May 2-4, and 1999 July
24-25. During these observations, the long-exposure seeing at 2.2 $\micron$ averaged $\sim$0$\farcs$6. 
The majority of the images were obtained for a proper motion study of the central
stellar cluster (Ghez et al. 1998, 2000) 
and were, therefore, centered roughly on the nominal position of Sgr A* (3$''$ Northwest of IRS 21); 
some images taken in 1998 August were centered on IRS 21. During most observations, additional frames
were also obtained to construct a larger mosaic, which included the three stars necessary to tie the near-infrared reference frame 
to both the radio and mid-infrared reference frames (IRS 7, IRS 10EE, and IRS 3, Menten et al. 1997). 
Each observation consisted of several sets of 
$\sim$100 short exposure (t$_{int}$ = 0.14 sec) frames, which freeze the distortion introduced by turbulence in the
Earth's atmosphere. Each of these speckle frames, which have a pixel scale of 0.0203$\pm$0.0005 arcsec/pixel and a 
corresponding field of view of 5$''\times$5$''$, contain high spatial resolution information that is recovered in post-processing. 

Diffraction-limited images were obtained from the speckle frames using the technique described in detail by Ghez et al. (1998).  In brief,   
the individual frames were processed in the standard
manner: bad-pixel-corrected, sky-subtracted, sub-pixelated by a factor of two and flat-fielded.  
The shift-and-add (SAA) images were generated by combining sets of 100 frames that have been shifted to align the brightest speckle of a bright point source located near the center of the 
field of view (either IRS 16C or IRS 33E for the frames centered on IRS 21).  
The resulting SAA point spread function (PSF) consists of a seeing halo and a diffraction limited core ($\theta_{diff}\simeq$0$\farcs$05)
containing $\sim$2-5\% of the total light (see Figure~\ref{cenpar2mic}). 

Additional K-band observations of IRS 21 were obtained with the W.M. Keck II Adaptive Optics (AO) system (Wizinowich et al. 2000) 
on the night of 1999 May 5. In these observations, USNO 0600-28579500, which has an R (0.7$\micron$)-band 
magnitude of 13.2 and is 33$\arcsec$ away from 
IRS 21, served as the natural guide star. The 256$\times$256 near-infrared camera, KCAM, used behind the 
AO system, has a plate scale of 0.0171$\pm$0.0004 arcsec/pixel. The plate scale was established based on the NIRC plate scale using the relative distances between 
IRS 16C, IRS 16SW, and IRS 16NE, as measured with KCAM and NIRC. The AO images, which consist of five-second exposures,
were calibrated and sub-pixelated in the same manner as the speckle frames. 
Five images, out of total of nine, with the best AO performance (core size less than 0$\farcs$2) were registered and averaged together.  
The PSF of the final image has a core-halo structure, with a 
core size of 0$\farcs$12 and 13\% of the light contained within the core. Compared to the AO performance on the 
Galactic Center earlier in the evening, as reported by Wizinowich et al. (2000), 
the AO performance during the IRS 21 observation was significantly degraded so that only half as much energy
was seen in the core. Nonetheless, compared to the typical PSF in the 
speckle images, the AO PSF for the IRS 21 observations has $\sim$5 times more energy in the core but a core
size that is about twice as broad. Neither the speckle nor AO data were flux calibrated. 

\subsection{Mid-Infrared Images} 

Mid-infrared images of the central parsec were obtained through narrow-band filters 
at 8.8, 12.5, 20.8, and 24.5 $\micron$ ($\Delta\lambda$ = 0.87, 1.16, 1.65, 0.76 $\micron$, respectively)  
with the MIRLIN mid-infrared camera (Ressler et al. 1994), which has a plate scale of 0.137$\pm$0.003 arcsec/pixel, based on 
the NIRC plate scale and the NIRC and MIRLIN measurements of the relative distances between IRS 7 and IRS 3, and a field of view 
of 17$''\times$17$''$ when mounted on the W. M. Keck II 10-meter telescope. 
The 8.8 and 24.5 $\micron$ data were obtained on 1998 March 13 \& 16 while the 12.5 and 20.8 $\micron$ data were collected on 1998 June 14-15. 
Images of the Galactic Center and a set of reference stars ($\alpha$ Sco, $\alpha$ Boo, $\alpha$ Lyr, $\beta$ Leo and $\sigma$ Sco)
were taken throughout the night, allowing for photometric\footnote{At 8.8 $\micron$, there were an insufficient
number of reference stars observed to establish an accurate zero point. We, therefore, used the Stolovy et al. (1996) measurements
of IRS 7 and IRS 3 to obtain a zero point for our 8.8 $\micron$ map.} and PSF calibration. 
For these observations, the chopping secondary was operated at a frequency of $\sim$4 Hz with an
amplitude of 15$''$ at a position angle of 315$\arcdeg$ in March and 30$''$ at 135$\arcdeg$, 150$\arcdeg$, and 170$\arcdeg$ in June. 
Likewise, the telescope was nodded at a frequency of 0.04 Hz with an amplitude of 60$''$ at a position angle of 270$\arcdeg$ in March
and 90$''$ at a position angle of 225$\arcdeg$ in June. Each chop image had a integration time of 10-70 milliseconds with 4-12 
coadds per image and 100-200 chops per frame resulting in a total on-source integration time ranging from 5 to 15 minutes over the wavelength range covered. 
The chop-nod pairs were double differenced, flat-fielded, bad pixel corrected, airmass calibrated, sub-pixelated by a factor of 6, 
registered and then rebinned by 2 pixels to create a $\sim$25$''\times$25$''$ mosaic 
of a region centered on Sgr A* and including IRS 21 (see Figure~\ref{mirn5}). The intrinsic fluxes for the photometric
standard stars are estimated using the convolution of the MIRLIN filter transmission and the quantum efficiency curves with the spectra 
of the photometric standard stars provided by Cohen et. al (1992).
Figure~\ref{irs21colmap} shows the color map, centered on IRS 21, constructed using the flux calibrated 12.5 and 20.8 $\micron$ images. 

\section{Results}

\subsection{Near-infrared Size of IRS 21 }

The remarkable extent of IRS 21 at 2.2 $\micron$ is evident in Figure~\ref{irs21con}, which contrasts the 2.2 $\micron$ image of 
IRS 21 and the point source, IRS 33E (Blum et al. 1996).
The image of IRS 21 shows no significant deviation from circular symmetry at this spatial resolution. Therefore, our analysis of its size is 
performed in one dimension by comparing the azimuthally-averaged intensity profile of IRS 21 with that of the PSF. 
Due to their relatively isolated positions within the 2.2 $\micron$ images, IRS 16NE is used as the PSF for all of the near-infrared
central cluster images,
while IRS 33E is used as the PSF for those images having IRS 21 centered in the field of view. Both IRS 16NE and IRS 33E 
appear to be unresolved stars at this resolution. Since the source density in this region
is large, a circular area with a radius of 0$\farcs$6 around any neighboring sources brighter than m$_{K}\sim$14 is 
excluded from the azimuthal averages. 
Each pixel included in the average is weighted by the number of frames that contributed to its value in the 
SAA processing or AO averaging. The background is subtracted on the basis of the median value of the radial profile between 1$\farcs$0 and 1$\farcs$1.
Figure~\ref{irs21kprof} shows examples of the resulting radial profiles for IRS 21 and a PSF. 

Once the radial intensity profiles are extracted, an intrinsic size is estimated by modeling the observed
IRS 21 profile as the convolution of the PSF profile with a
Gaussian function. Figure~\ref{irs21kprof} also shows the best fitting model profile, from which we use the Gaussian 
HWHM to assign an intrinsic radius of IRS 21 for each image.
Table 1 lists the average radius and standard deviation from all epochs of 2.2 $\micron$ data. 
The 2.2 $\micron$ radius derived from the SAA images appears to have remained constant from 1995 to 1999, with a weighted average and 
uncertainty of 725$\pm$40 AU for all epochs of data. Although the AO result is somewhat smaller, 570$\pm$40 AU, this is most likely
due to a mismatch between a Gaussian and the true intrinsic shape. Nonetheless, this modeling allows us to assign an
effective size, which we take to be the average of these two values, 650$\pm$80 AU, assigning the half-range as our uncertainty.   

\subsection{Mid-infrared Size, Identification and Fluxes of IRS 21} 

Figure~\ref{irs21midcon} (column 1) shows the 2.2 $\micron$ position of IRS 21 
in the 8.8, 12.5, 20.8, 24.5 $\micron$ maps of IRS 21, based on the near- and 
mid-infrared positions of IRS 7 and IRS 3.   Within the 
2$''\times$2$''$  region displayed, IRS 21's mid-infrared emission peaks within
0$\farcs$1$\pm$0$\farcs$1 of its 2.2 $\micron$ position.
A second mid-infrared peak is also detected 0$\farcs$5 Northwest of IRS 21; however, the color map provided in Figure~\ref{irs21colmap} 
shows that only the source associated with IRS 21 has a color that is distinguishable from the background. 
This suggests that the neighboring 
source is an externally heated region of increased dust density as was first proposed to be the case for 
IRS 21 (Gezari et al. 1985).

The structure of the diffuse background emission from the Northern Arm and the close proximity of the neighboring source 
(see Figures~\ref{mirn5} and~\ref{irs21midcon}) complicates the analysis of the extent of IRS 21 at 
mid-infrared wavelengths compared to that carried out at 2.2 $\micron$. 
In order to separate IRS 21 from the background emission of the Northern Arm, which has a ridge-like structure, 
and neighboring source, we apply an algorithm similar to CLEAN (Hogbom 1974), 
which removes IRS 21 from the image, leaving only the diffuse emission. Since IRS 21 lies on the edge of the Northern Arm, 
we create a first-order approximation of the surrounding background emission by fitting an inclined plane
to the region within an annulus, 1$\farcs$0 in outer radius and 0$\farcs$2 wide, around the position of IRS 21. 
To avoid the flux from the discrete neighboring source when fitting the plane, the area within 0$\farcs$4 of the source ($\sim$25\% of the annulus) was 
excluded from the fit. After subtracting the inclined plane from the image, the CLEAN algorithm iteratively subtracts a scaled ($\gamma$=0.5) PSF from the 
position of the peak flux within a 1$\farcs$0 cleaning radius of the position of IRS 21. Here, where a 2-D estimate of the PSF is necessary, the independently
observed photometric standards were used as estimates of the PSF.     
The progress of the cleaning algorithm is monitored by calculating the average flux within the
cleaning radius and is halted when this flux falls below zero. Two maps are created by this procedure, a residual map and a delta map (see Figure~\ref{irs21midcon}).
The residual map (not shown) contains only the background emission from the Northern Arm and the delta map contains IRS 21 and
a small contaminating contribution from the neighboring source. In order to isolate IRS 21, those clean components within a radius of 0$\farcs$4 from the 
position of the neighboring source as well as those found within a radial interval of 0$\farcs$2 about the 1$\farcs$0 cleaning radius are reconvolved with the scaled PSF 
and added back into the residual map. Since the resulting residual map should represent the diffuse emission of the Northern Arm and the neighboring source, 
subtracting it from the original image leaves only the flux from IRS 21 (depicted in column 2 of Figure~\ref{irs21midcon}). 

The radial extent of IRS 21 at mid-infrared wavelengths is estimated from  
azimuthally-averaged radial intensity profiles of the background-subtracted images with the same 
method used to analyze the 2.2 $\micron$ profiles, with one additional step. An estimate of the 
systematic uncertainty from the imposed size of the cleaning radius is obtained by varying the cleaning radius from 0$\farcs$9 to 
1$\farcs$1. This is added to the statistical uncertainty, which is estimated using the RMS of the results from the analysis on 
the three sub-sets of the MIRLIN images. The statistical uncertainties are the dominant source of error ($\S$ 3.1). 
The radial extent of IRS 21 is estimated at each wavelength  (see Figure~\ref{midir_prof}) using 
the following two sources for the PSF: 
(1) the photometric standard star and (2) IRS 7, which is in the same field of view as IRS 21 but for which IRS 3 was masked out in order to create
an accurate 1-D PSF estimate.  
At all wavelengths except 8.8 $\micron$, the separately observed PSF star and IRS 7 have a FWHM that corresponds roughly to the diffraction
limit and therefore yield comparable results. The values quoted in Table 2 at these longer wavelengths are based on the separately
observed PSF. Between 12.4 and 24.5 $\micron$ the size estimates 
are not significantly different as a function of wavelength and produce a mean radius for IRS 21 of 1600$\pm$200 AU. 
At 8.8 $\micron$, the FWHM of IRS 7 and IRS 3 are both twice as large as that of the separately observed
point source, $\beta$ Leo (see Table 2), suggesting a change in atmospheric conditions (i.e., the IRS 21 images at 8.8 $\micron$ are seeing limited). 
Since the radial profile of the simultaneously observed IRS 7 is comparable in size to 
IRS 21 at 8.8 $\micron$, we cannot estimate a size for IRS 21 at this wavelength and do not include this radial profile 
in our modeling presented in $\S$4. A 3$\sigma$ upper limit to the 8.8 $\micron$ size of IRS 21 ($<$1830 AU) is assigned.
This is based on the FWHM of the Gaussian which, when convolved with the radial profile of IRS 7, results in a reduced 
chi squared value of 3 between the model and observed radial profile.  

Table 2 also lists the mid-infrared fluxes measured for IRS 21 by comparing the total counts within a 1$\farcs$0 radius circular
aperture in the IRS 21 image with and without the background subtracted (column 2 in Figure~\ref{irs21midcon}) 
to the counts within the same aperture around a photometric standard star.
We estimate that IRS 21 is 31, 15, 6 and 6\% above the background
value in the original images at 8.8, 12.5, 20.8 and 24.5 $\micron$, respectively.
The uncertainties for the fluxes given in Table 2 are the sum of the systematic and statistical uncertainties as estimated by the
same method used for the mid-infrared sizes. Compared to the previously reported 8.7 $\micron$ flux value of 5.6$\pm$2.0 Jy
by Stolovy et al. (1996) using a 2'' circular aperture and no background removed, our values are lower due to a smaller aperture size. 
Using the same 2'' circular aperture and not removing the Northern Arm contribution, we obtain a comparable value of 
3.6$\pm$0.1 Jy at 8.8 $\micron$. No other mid-infrared photometry for IRS 21 has been reported in the literature. 

\section{Discussion} 

\subsection{Origin of the Near- and Mid-infrared Dust Cloud}

The immense size of IRS 21 at 2.2 $\micron$ ($\sim$650 AU) and in the mid-infrared ($\sim$1600 AU) is well beyond 
that expected for a stellar photosphere by over two orders of magnitude. 
For comparison, the radius of the largest stellar photosphere, an M super-giant, is roughly 4 AU (Drilling \& Landolt 2000). 
We are, therefore, most likely detecting an extended distribution of dust around IRS 21. The color temperature
of IRS 21 from our MIRLIN data (see Figure~\ref{irs21colmap}) shows that it is hotter than the surrounding Northern Arm dust, 
suggesting that it is self-luminous and, therefore, has a central radiative source (Gezari 1992).
If the dust is intrinsic to the heating source, it could be in the form of an inflow, as in
the case of a protostar, or an outflow of material, as in the case
of, for example, a Wolf-Rayet star, a luminous blue variable (LBV), or an M supergiant.
Alternatively, the dust could be associated with the infalling Northern Arm, either passively
heated by a central radiative source and/or dynamically perturbed by the stellar winds of IRS 21. 
In either case, the Roche limit for a star of mass, M, at the projected distance of IRS 21 from Sgr A* is  
480 AU (M/10 M$_{\odot}$)$^{1/3}$, which suggests that 
the current dust distribution is transitory rather than being a long-lived 
configuration or a static shell. This implies that the dust is unlikely to be part of an inflow
(unless the projection effects are quite strong), adding to the list of difficulties for the young stellar object
hypothesis introduced in \S 1. 

\subsection{The Spectral Energy Distribution}

Construction of the spectral energy distribution (SED) for IRS 21 provides additional insight into the nature of the source. 
Figure~\ref{all} combines our mid-infrared flux densities for IRS 21 and the Northern Arm (see Table 2) along with others collected 
from the literature (references given in the Figure~\ref{all} caption). All of the 
photometry from the literature is calibrated using zero points from Cohen et al. (1992) and dereddened assuming a visual 
extinction of A$_V$=30 and an ISM extinction law
based on observations of extinction at the Galactic Center (see Moneti et al. 2001 and
references within). The shape of the dereddened SED suggests there are two distinct components:  
one in the near-infrared responsible for the flux between 2-4 $\micron$, and a separate, mid-infrared component. 
The mid-infrared component is assumed to be thermal emission from dust heated by both the central source and through the trapping of external 
Lyman $\alpha$ photons. The latter is necessary to 
achieve the grain temperatures required to model the mid-infrared component of the SED (see Rieke, Rieke, \& Paul 1989 and references within).
We ascribe the near-infrared component to dust scattering as well as extincted light from the central source. 
Scattering is further supported by the large 2.2 $\micron$ polarization (Ott et al. 1999). This implies
an asymmetry in the dust distribution that is below our angular resolution. 

To utilize both the spectral and spatial information, we have created a simple 2-D radiative transfer code that simultaneously
fits the 2.2 to 24.5 $\micron$ flux densities and the radial profiles including both the flux from IRS 21 and the Northern Arm (see Figure~\ref{all}).  
The basis for the model is a central blackbody embedded in a distribution of
gas and dust, which singly scatters or absorbs light from the central source. The central radiative source is described by two free parameters: its
radius, R$_{cen}$, and temperature, T$_{cen}$. In addition to the internal heating from the central
blackbody, an external heating contribution is added in the form of Lyman $\alpha$ trapping\footnote{Lyman $\alpha$ trapping is a function of the gas density, n$_H$, which is estimated
through the dust density and the gas to dust mass ratio. It is also a function of $z_{L \alpha}$ which is the fraction of recombinations resulting
in the production of a Ly $\alpha$ photon. This parameter is dependent on the electron temperature, which is assumed 
to be 7,000 K (see Roberts \& Goss 1993) resulting in a recombination fraction of 0.7 (Spitzer 1978)}.
For simplicity, we assume that the same, single population of dust grains is responsible for both the thermal and scattering component. 
This model does not account for grain heating due to light scattered within the dust.
Fixed in this model are (1) the gas to dust mass ratio of 100:1 (2) the dust's emission and scattering
coefficients, which are assumed to be those derived for silicate grains\footnote{The mid- to far-infrared 
spectrum of the Northern Arm has been previously modeled with optically thin silicate emission by Chan et al. (1997). Furthermore,
graphite grains produce significantly worse fits for the SED models.}
by Laor \& Draine (1993) and (3) the distribution
of dust grain radii, n(a), which is assumed to depend on dust size, a, as n(a)$\propto$a$^{-3.5}$da from 0.02 to 0.25 $\micron$ (Mathis et al. 1977, MRN).
A number of possible dust density profiles are considered, including  a constant dust density model, which 
represents passive heating of the Northern Arm ($\S$4.3), a 1/r$^2$ dust density model appropriate for an
outflow or inflow of material assuming a constant velocity ($\S$4.4),
and a ``bow shock'' dust density model, which would be expected if the Northern Arm were interacting with stellar winds from the
central radiative source ($\S$4.5). 

Model fitting proceeds by minimizing the reduced $\chi^2$ value for both the SED and four (2.2, 12.5, 20.8 and 24.5 $\micron$) 
radial profiles\footnote{The near- and mid-infrared radial profiles are rebinned to match the original plate scales prior to
convolution with the radial profile of the PSF.}  
simultaneously. We've chosen to equally weight the near-infrared photometry, mid-infrared photometry, the near-infrared profile,
and the mid-infrared profiles ($\tilde{\chi}^2_{tot}$=$\frac{1}{2}\tilde{\chi}^2_{SED}$+$\frac{1}{4}\tilde{\chi}^2_{2.2profile}$ 
+$\frac{1}{12}\tilde{\chi}^2_{12.5profile}$+$\frac{1}{12}\tilde{\chi}^2_{20.8profile}$+$\frac{1}{12}\tilde{\chi}^2_{24.5profile}$).
The weighted reduced chi squared provided above is minimized to find the best fitting value for each free parameter. 
In order to avoid local minima, this is done using 100 iterations of the Powell minimization algorithm with randomly chosen sets 
of initial parameters (Press et al. 1992). The range of the initial values is set to cover a reasonable portion of acceptable values 
for the physical parameters. The uncertainty in each fitted parameter is estimated using the amount of variation around the best-fit value that
produces a deviation in $\chi^2_{tot}$ representing a 3$\sigma$ confidence level given the number of free parameters 
in each model (Press et al. 1992). 

\subsection{Constant Dust Distribution} 

As a first guess to the distribution of the dust around IRS 21, we place the central heating source 
within a constant density of material. 
The only free parameters for the constant dust density model are the size and temperature of the central source, 
the radius of the inner edge of the dust, r$_i$, and the dust mass density, $\rho_d$.
Table 3 gives the parameters of the best fitting model ($\tilde{\chi}^2$=800). This model did not provide an 
acceptable fit to the radial profiles as it produces too much flux in the wings of the mid-infrared profiles (Figure~\ref{all}, top row). 
This suggests that the dust density falls off at some distance from the inner edge of the dust envelope as might
occur if the surrounding dust were undergoing some type of dynamical process, either intrinsic mass in- or out-flow or interaction
with the Northern Arm. Based on this analysis, we rule out the model in which the infrared extended
emission arises from dust in the Northern Arm that is passively interacting with an embedded central source.

\subsection{1/r$^2$ Dust Distribution}

For a constant mass flow rate, a spherical outflow or inflow will produce a 1/r$^2$ density distribution.
The free parameters for the dust distribution in this model 
include the radius of the inner and outer edge of the dust, r$_i$ and r$_o$, the dust mass density at the inner edge, 
$\rho_d$, and the dust density of the surrounding Northern Arm, $\rho_{NA}$. This results in a total of six free parameters, 
including the temperature and radius of the central source. Table 3 gives the parameters and errors for the best fitting
model ($\tilde{\chi}^2$=80) which is also plotted in Figure~\ref{all} (middle row). 
The size and temperature of the central source are suggestive of an embedded, optically thick dust shell.
The inner and outer extent of the dust are consistent with the near- and mid-infrared Gaussian sizes listed in Tables 1 and 2. 
The luminosity of the central object (10$^4$ L$_{\sun}$) allows for several possibilities 
for its identity: a massive main sequence or post-main sequence early-type 
star (M$>$25 M$_{\sun}$), a massive AGB star, or a Wolf-Rayet star (see Table 4).

Although the 1/r$^2$ model yields a somewhat reasonable fit to the SED and radial profiles, given the simplicity of the model, 
there are arguments against the constant mass flow model. 
Using the derived gas density and typical outflow velocities of luminous stars in Table 4, 
we estimate mass-loss rates (column 5) which are at least an order of magnitude above the values expected for all of the stellar candidates.
Furthermore, no known intrinsic outflow source has an observed infrared radius due to optically thin emission comparable to that observed for IRS 21
($\sim$650 AU at 2.2 $\micron$ and $\sim$1640 AU at mid-infrared wavelengths). 
Examples of large resolved outflow sources at 2.2 $\micron$ include the spiral dust shell 
around the binary Wolf-Rayet star, WR 104, and the dust shell around the Wolf-Rayet star, Ve 2-45, which have diameters of 160 AU and 
70 AU, respectively (Tuthill, Monnier, \& Danchi 1999; Danks et al. 1983).  
The well-studied carbon star IRC +10216, whose dust shell is optically thick at 2.2 $\micron$, has an estimated radius of 
$\sim$50 AU (Rowan-Robinson \& Harris 1983). 
There are protostellar objects with 2.2 $\micron$ radii on the order of a few hundred AU as measured using speckle
interferometry (Howell, McCarthy \& Low 1981), but as discussed in $\S$1 and $\S$4.1 there are difficulties with the protostellar hypothesis. 
These arguments suggest that the constant mass flow model alone is not a suitable model for IRS 21.

\subsection{Bow Shock Dust Distribution} 

A model that allows for a condensed dust distribution without requiring a large intrinsic dust outflow, is a
``bow shock'' model. The bow shock dust density profile is modeled as a spherical shell, with a constant
dust density from an inner radius of r$_i$ to an outer radius of $r_o$. While a bow shock does not
wrap completely around the star which produces it, we judge that the spherical approximation will
provide a reasonable estimation of the radial profiles and the photometry in the case of a relatively face-on bow shock. 
Table 3 gives the parameters and errors for the best fitting model ($\tilde{\chi}^2$=30), which is shown in the
bottom row of Figure~\ref{all}. The gas density of 
the Northern Arm is consistent with the expected value of 10$^5$ cm$^{-3}$ given by Genzel et al. (1998).
We argue below that the bow shock model is a physically sensible explanation for the optically resolved structure\footnote{Since this simple radiative transfer
model contains a number of assumed, but undetermined parameters, we do not report formal uncertainties. A better fit to the 2.2 $\micron$ radial profile can be achieved using a 
larger upper limit to the grain size. The addition of larger grains to the grain population
increases the scattering coefficient such that the 2.2 $\micron$ flux contains a greater fraction
of scattered light. This results in a 2.2 $\micron$ radial profile with a broader core, thus providing
a better fit to the observed near-infrared radial profile. We, however, do not feel that this is evidence for large grains, given the simplicity of the model.}.

The temperature and radius of the {\it central source} as well as the lack of any observed photospheric lines suggest that it is
an optically thick dust shell from a 1.3$\times$10$^4$L$_{\sun}$ mass losing star. This central source 
is surrounded by the optically thin dust that has been modeled here as a bow shock to explain the resolved emission. 
However, the luminous sources listed in Table 4  are not equally capable of producing bow shocks 
from interactions with the Northern Arm at a distance of $\sim$1000 AU.  We have estimated the stand-off distance, $l$, which is
defined as the distance at which the ram pressure from the intrinsic stellar wind equals that of the motion 
of the material flowing along the Northern Arm.
$l$ = 1.74$\times$10$^{19}$$\dot{m_*}$$^{1/2}$v$_w^{1/2}$v$_*^{-1}$$\mu_H^{1/2}$n$_H^{-1/2}$ cm, where $\dot{m_*}$ is the stellar mass loss 
rate in units of 10$^{-6}$ M$_{\sun}$ yr$^{-1}$, v$_w$ is the 
stellar wind velocity in units of 10$^8$ cm s$^{-1}$, v$_*$ is the motion of the star through the
Northern Arm in units of 10$^6$ cm s$^{-1}$, $\mu_H$ is the mean molecular weight and, n$_H$ 
is the gas density of the Northern Arm in units of cm s$^{-1}$ (Weaver et al. 1977). 
We assume a relative velocity of $\sim$100 km s$^{-1}$ between
IRS 21 and the Northern Arm material (Herbst, Beckwith, \& Shure 1993), the dust density of the Northern Arm 
derived from the bow shock radiative transfer model, $\rho_{NA}$, and the mass loss rates and stellar wind
velocities of each candidate as found in the literature (see Table 4). The resulting bow shock distances are given in column 6 of Table 4. 
While there are many stars with luminosities near that of IRS 21, they have a range of stellar wind velocities and mass
loss rates affecting their expected stand-off distance. Based on comparison of the derived stand-off distances and the observed size of IRS 21, 
we dismiss the possibility that IRS 21's central source is a giant star, like IRS 7, and suggest that it is possibly a massive star in a more evolved
state, such as a Wolf-Rayet or AGB star. 

The bow shock scenario is appealing because it explains the association of 
IRS 21 and the Northern Arm without requiring that the central source be formed within
the Northern Arm before or during the infall toward Sgr A$^*$.
A bow shock instead requires the much more plausible circumstance that IRS 21 is simply passing through the Northern Arm, resulting in
an interaction with its intrinsic stellar winds. The apparent absence of the
characteristic horse-shoe geometry indicative of a bow shock, can be attributed to the orientation 
(if the vector representing the relative velocity of IRS 21 and the Northern Arm is oriented predominantly toward or away us) 
or to the depth at which the central heating source is embedded 
within the Northern Arm. Scattered light by an asymmetric dust distribution, however, would 
explain the large observed K-band polarization (Ott et al. 1999).
Finally, the bow shock model receives support from the recent 2.2 $\micron$ Gemini AO observations
of a clear bow shock structure surrounding the possibly analogous Northern Arm source, IRS 8, located 32$\arcsec$ north of IRS 21 (Rigaut 2001).

\section{Summary and Conclusions}

The near- and mid-infrared light from IRS 21, a luminous source at the Galactic Center, has been spatially resolved, giving a remarkably large
radius of $\sim$1000 AU. Using this size information along with the 1-25 $\micron$ photometry, 
we suggest that IRS 21 is embedded in the dense ``Northern Arm'' component
of Sgr A West, and that the obscured starlight is diffusing out through the optically thin dust along this feature. 
The near-infrared light is scattered and extincted light, which probably originated in a dust photosphere overlying the 
central star, and the mid-infrared light is thermally re-radiated dust emission from the optically thin shell. 
Furthermore, our SED and radial profile modeling, 
along with high-resolution imaging of the similar Northern Arm mid-infrared source, IRS 8, suggest that the dust 
surrounding IRS 21 is in the form of a bow shock produced as the Northern Arm encounters the winds from 
the central star. The bow shock scenario removes the necessity that IRS 21 be a young source that formed within the Northern Arm 
during or prior to its infall toward Sgr A$^*$, and suggests that the central stellar source emits strong winds, accounting for the observed size. 
The featureless 2.2 $\micron$ spectrum suggests that the central source is an optically thick dust shell surrounding
a mass-losing source such as a dusty WC9 Wolf-Rayet star. 
IRS 21 may serve as a prototype for the other luminous sources lying along the Northern Arm, including
IRS 1W, 2, 10W, and 5. Considering the large population of windy He I stars within the 
central parsec (Sellgren et al. 1990, Genzel et al. 1996), it is plausible that these Northern Arm 
sources are also producing bowshocks like IRS 21 and IRS 8. 

\begin{acknowledgements}

We would like to thank Tetsuya Nagata and Mike Jura for their helpful suggestions on the manuscript and Dan Gezari for 
providing us with unpublished mid-infrared photometry, we used
these fluxes for initial SED fitting prior to the completion of our own 
mid-infrared analysis.
The speckle imaging and mid-infrared work
were supported through NSF grants No. AST-9457458 and AST-9988397
and the adaptive optics analysis was funded 
by the NSF Science and Technology Center for Adaptive Optics, 
managed by the University of California at Santa Cruz under cooperative 
agreement No. AST-9876783. 
Portions of this work were carried out at the
Jet Propulsion Laboratory, California Institute of Technology, under
and agreement with NASA. Development and operation of MIRLIN is supported by
the JPL Director's Discretionary Fund and by NASA's Office of
Space Science.
\end{acknowledgements}

\newpage
 
\figcaption[cenpar2mic.ps]{A $\sim$3$''\times$3$''$ section of a typical 2.2 $\micron$ SAA image containing IRS 21 (t$_{int}$=26 seconds). This sub-image, which is composed
of data taken on 1999 May 15, shows the contrast between the point source, IRS 33E, and the resolved structure of IRS 21. North is up and east is to the left. 
 \label{cenpar2mic}}

\figcaption[gc_n5.ps]{Mosaic of a $\sim$25$''\times$25$''$ region of the Galactic Center at 12.5 $\micron$. The Northern Arm sources
(IRS 21, 1, 2, 10, and 5), IRS 3, and the red supergiant, IRS 7 are identified. North is up and east is to the left. 
 \label{mirn5}}

\figcaption[irs21colmap.ps]{A 12.5-20.8 $\micron$ color map grey scale image of the 1$''\times$1$''$ region around IRS 21 along with contours of the
same region at 12.5 $\micron$. The hottest region (dark) in the linear grey scale image is $\sim$ 250 K while the coolest regions (light) are
$\sim$ 200 K. The contours plotted represent 90-10\% of the peak flux value in 10\% increments.
The source 0$\farcs$5 Northwest of IRS 21, discussed in $\S$3.2, is absent from the color map suggesting it
is an externally heated region of higher dust density as was originally suggested for IRS 21 by Gezari (1985). The cross represents the 
2.2 $\micron$ position and uncertainty of IRS 21. Before the two images were divided, the 12.5 $\micron$ image was convolved 
with a 0$\farcs$4 FWHM Gaussian beam to match the angular resolution of the two images. North is up and east is to the left. 
\label{irs21colmap}}

\figcaption[irs21con.ps]{Contours plots of IRS 21 (left) and its neighboring point source, IRS 33E (right). 
The contours plotted for IRS 21 and IRS 33E represent 90-50\% of the peak value in 10\% increments and clearly show that IRS 21 is significantly resolved. 
\label{irs21con}}

\figcaption[irs21kprof.ps]{Keck AO (left) and SAA (right) radial profiles of IRS 21 (filled circles) 
and a PSF star (empty circles) along with the profile of the 
convolution of the PSF with a Gaussian (solid line). This modeling suggests a 2.2 $\micron$ intrinsic radius of $\sim$650 AU. 
The error bars represent the standard deviation of the mean intensity at each radius. 
\label{irs21kprof}}

\figcaption[irs21midcon.ps]{Plots of a 3''$\times$3'' region around IRS 21 showing various aspects of the ``cleaning'' process (see $\S$ 3.2). 
The original map or ``dirty map''and the PSF or ``dirty beam'' are shown in columns 1 and 3. 
Column 4 is the delta map depicting the positions and the intensity of the peaks subtracted
during the cleaning process including the circles which mark the region around the neighboring source and the annulus around IRS 21 within which the 
associated flux was added to the background image. Column 2 displays the final image of IRS 21 with the background removed. 
The contours plotted represent 90-10\% of the peak value in 10\% increments. The cross represents the 
position and uncertainty of IRS 21 in the 2.2 $\micron$ image. 
\label{irs21midcon}}

\figcaption[midir_prof.ps]{ a) Radial profiles of the background subtracted images of IRS 21 (filled circles) and a 
PSF star or IRS 7 (8.8 $\micron$ only) (empty circles) along with the profile of the 
convolution of the PSF with a Gaussian (solid line) at 12.5, 20.8 and 24.5 $\micron$. This model suggests a mid-infrared radius of 
$\sim$1640 AU. \label{midir_prof}}

\figcaption[all.ps]{{\it Left column} - De-reddened photometry and the best fitting constant (top), 1/r$^2$ (middle), and bow-shock (bottom) radiative transfer models 
for IRS 21 showing the two components of the SED - the near-infrared scattered light (dot-dashed line) and mid-infrared re-emitted light (dashed line) - 
as well as the sum of the two components (solid line). References --
J (1.25 $\micron$) - Blum et al. 1996;
  (1.45  $\micron$) - Stolovy private com.;
H (1.6 $\micron$) - Blum et al. 1996;
  (1.9 $\micron$) - Stolovy private com.;
K (2.2 $\micron$) - Blum et al. 1996, DePoy \& Sharp 1991, Simon et al. 1990 ;
L (3.5 $\micron$) -  Tollestrup, Capps \& Becklin 1989 ;
L' (3.8 $\micron$) - Simons \& Becklin 1996 ; 
8.8-24.5 $\micron$ - this work.
{\it Right column} - Radial profiles of IRS 21 (filled circles) along with the profile of the 
convolution of the PSF with the radial profile created by the constant (top), 1/r$^2$ (middle), and bow-shock (bottom) radiative 
transfer models at 2.2, 12.5, 20.8 and 24.5 $\micron$ (solid line). The mid-infrared photometry and radial profiles include a constant
background term, which is modeled to arise from Lyman $\alpha$ trapping in all cases. \label{all}}

\clearpage

\begin{figure}[h]
\epsscale{1.0}
\plotone{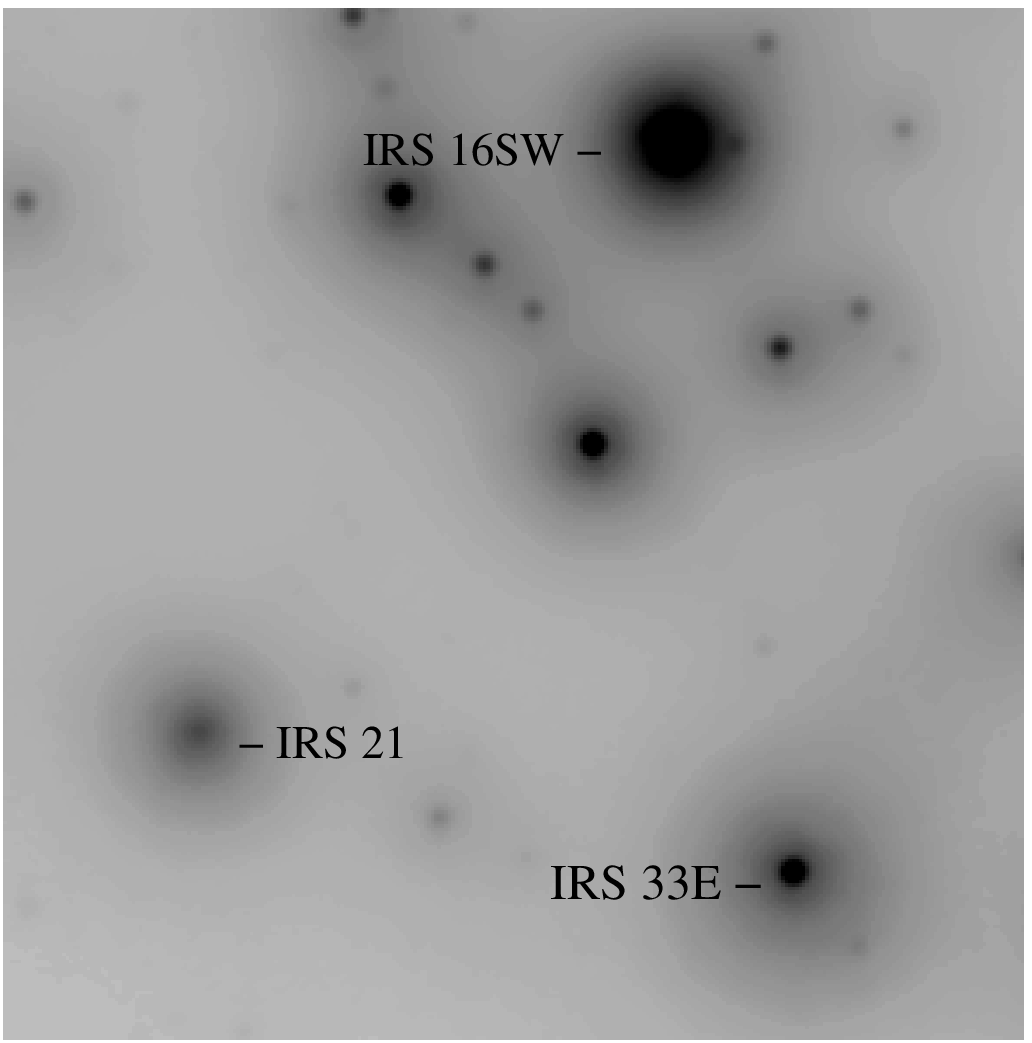} 
\end{figure}

\clearpage

\begin{figure}[h]
\epsscale{1.0}
\plotone{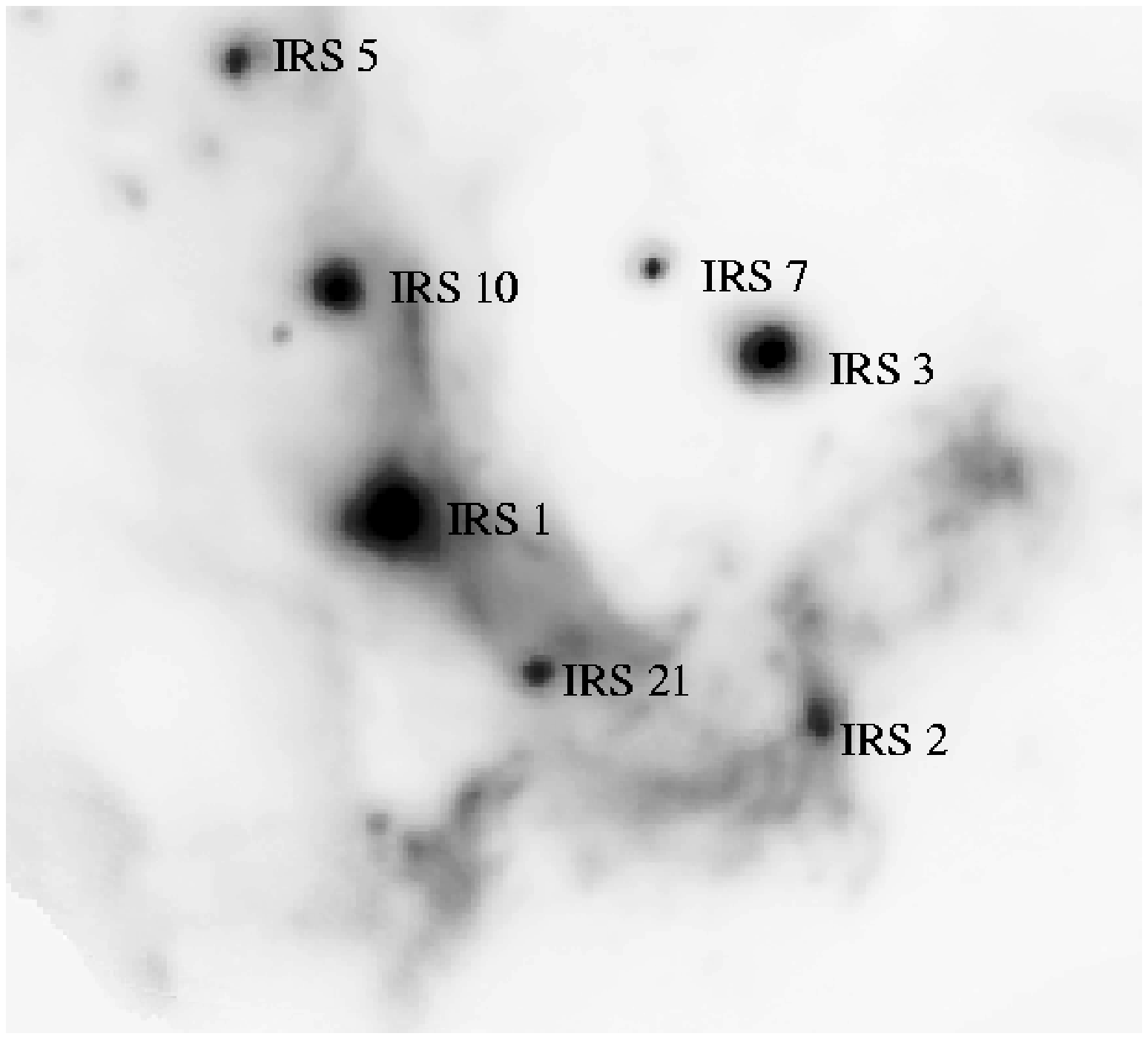}
\end{figure}

\clearpage

\begin{figure}[h]
\epsscale{1.0}
\plotone{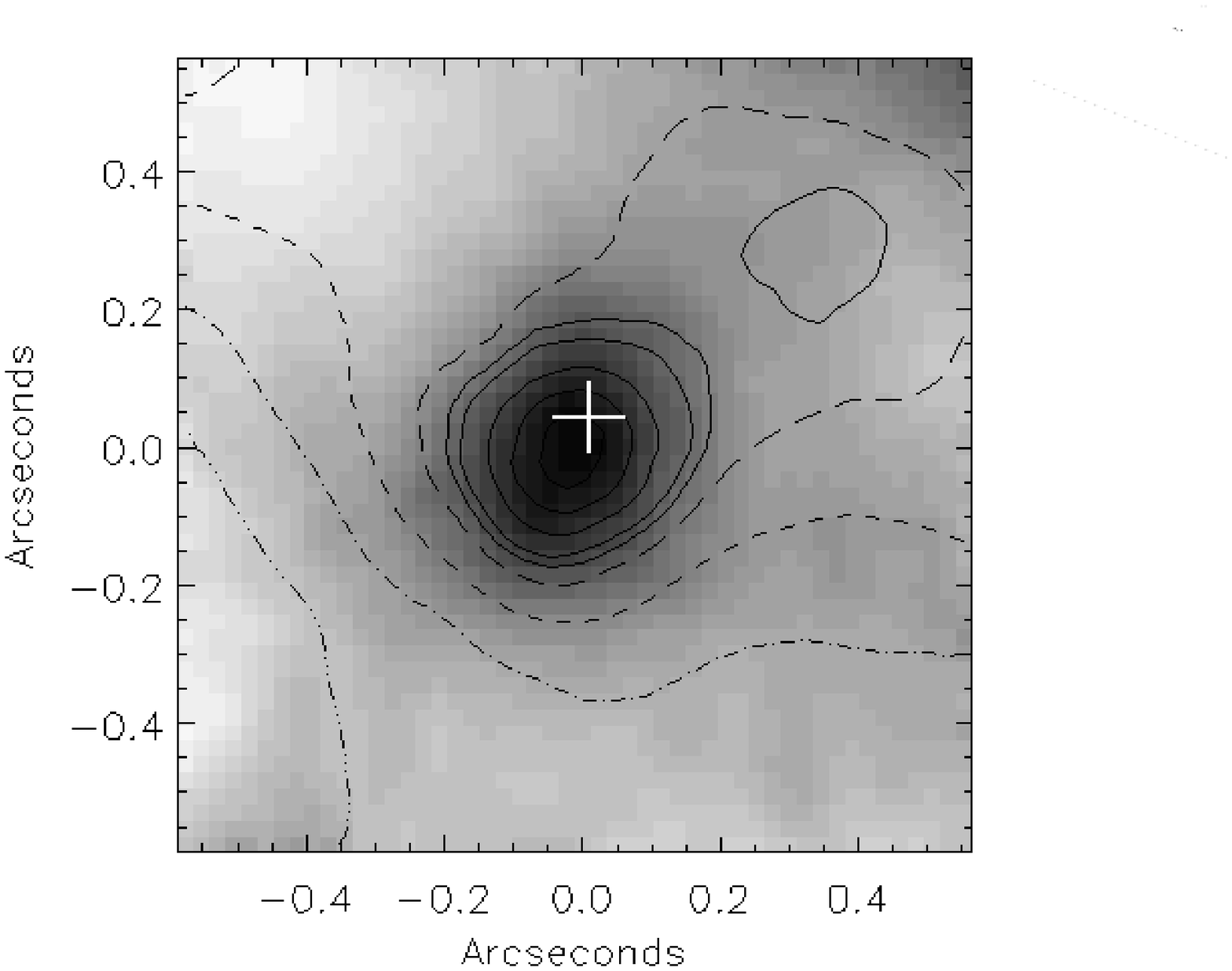}
\end{figure}

\clearpage

\begin{figure}[h]
\epsscale{0.5}
\plotone{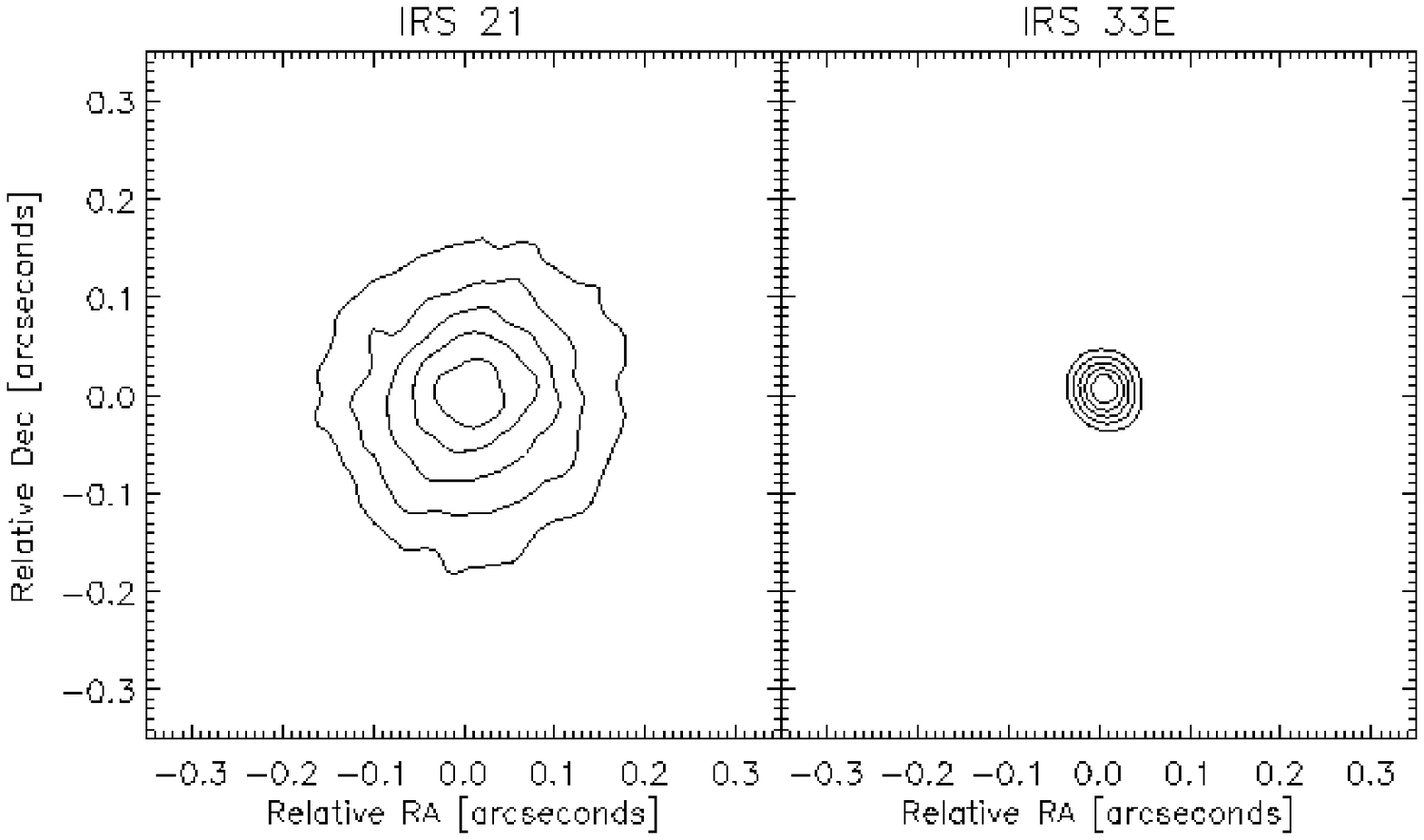}
\end{figure}

\clearpage

\begin{figure}[h]
\epsscale{1.0}
\plotone{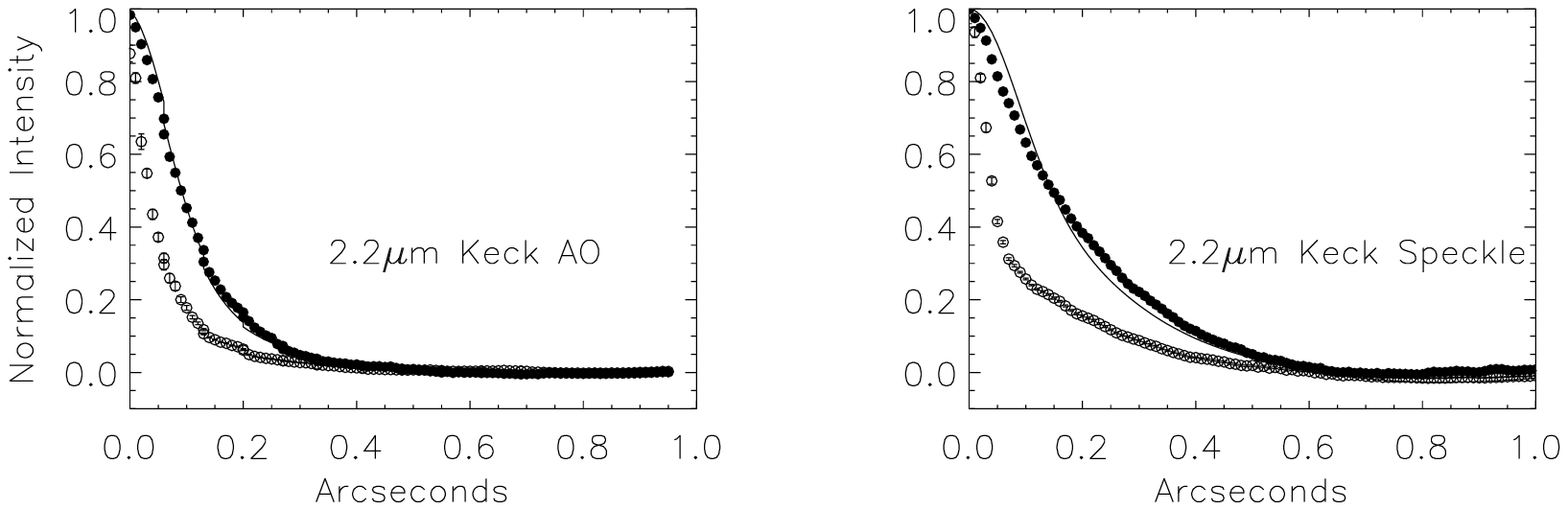}
\end{figure}

\clearpage

\begin{figure}[h]
\epsscale{1.0}
\plotone{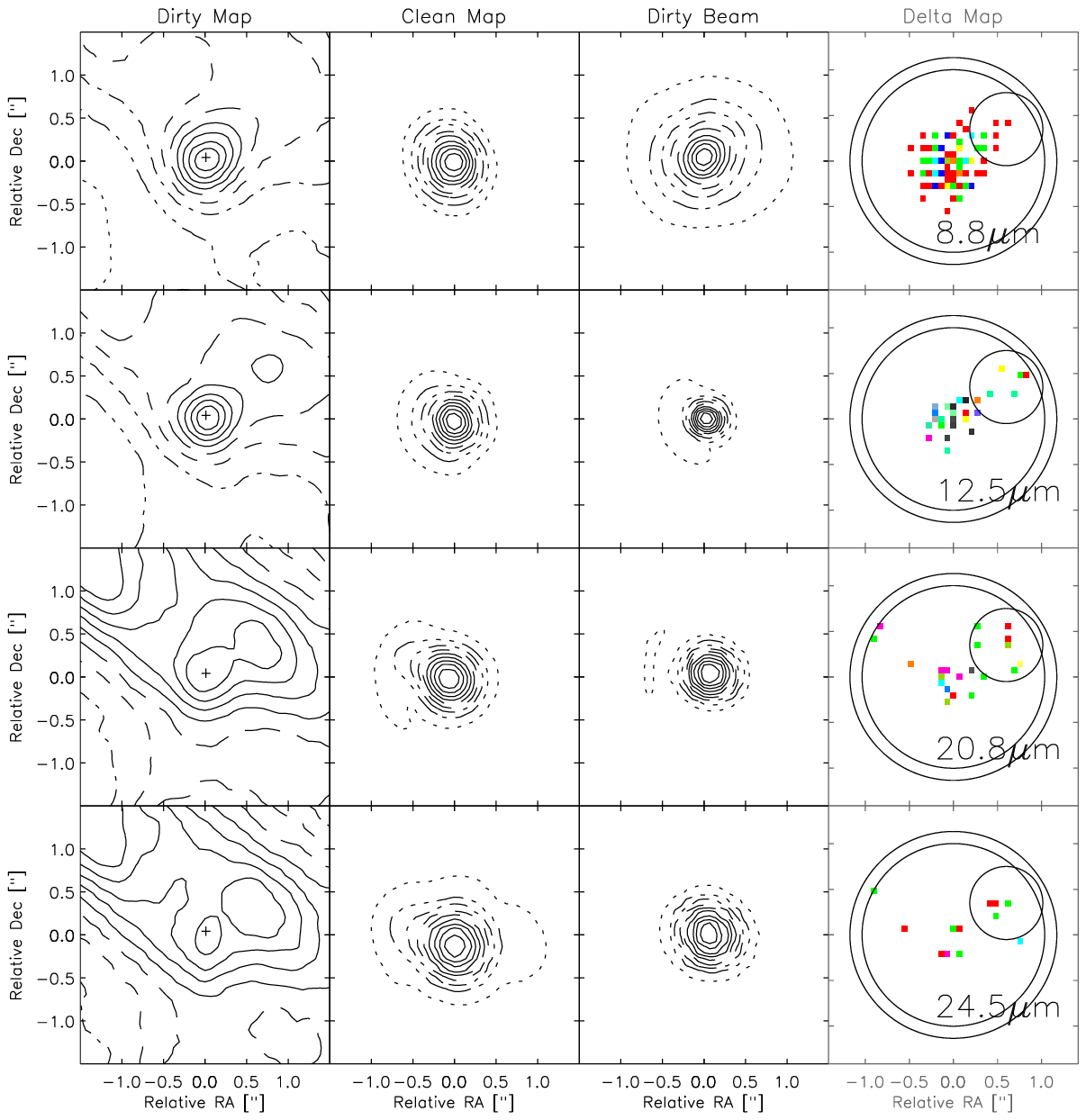}
\end{figure}

\clearpage

\begin{figure}[h]
\epsscale{0.75}
\plotone{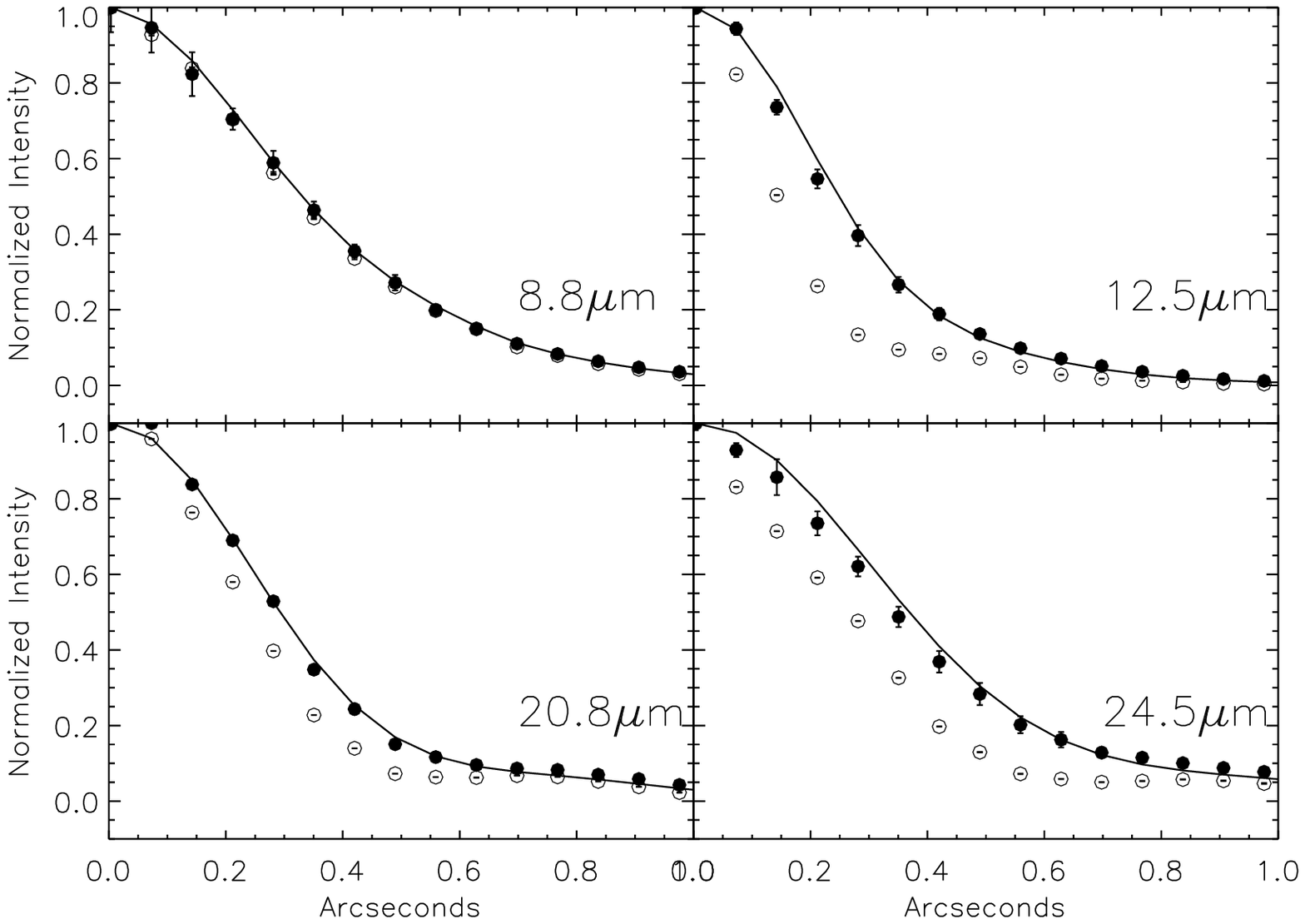}
\end{figure}

\clearpage

\begin{figure}[h]
\epsscale{1.0}
\plotone{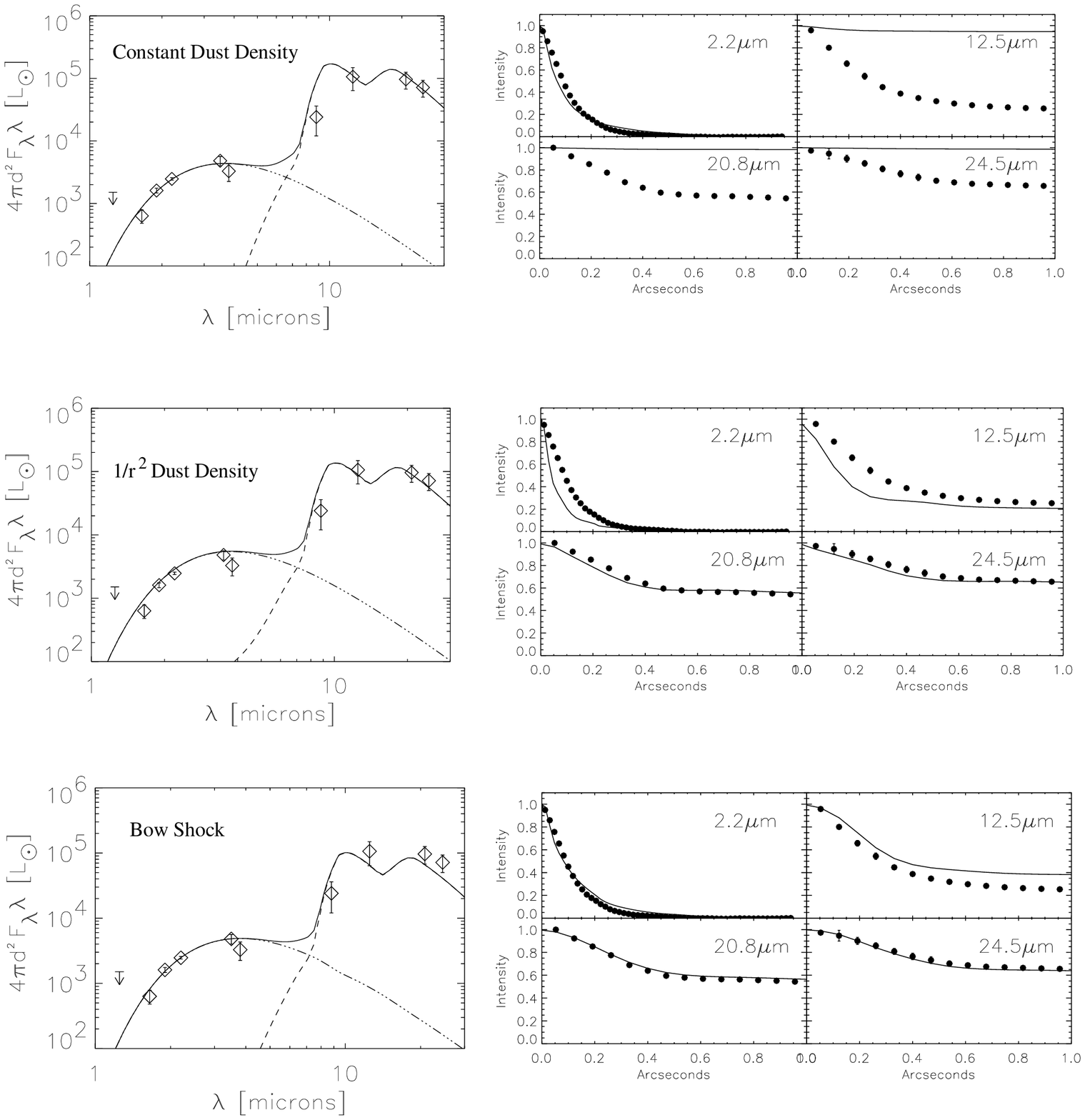}
\end{figure}

\clearpage 

\begin{deluxetable}{cccc}
\tablecaption{2.2 $\micron$ Radius of IRS 21}  
\tablehead{
\colhead{Date} & \colhead{Technique} & \colhead{\# of frames} & \colhead{Radius$^a$ [AU]} }
\startdata
1995 June  & Speckle  & 800   &   680$\pm$70  \nl
1996 June  & Speckle & 600   &   760$\pm$170  \nl
1997 May   & Speckle & 900   &   750$\pm$130 \nl
1998 April & Speckle & 588   &   790$\pm$115  \nl
1998 May   & Speckle & 980   &   900$\pm$90  \nl
1998 August & Speckle & 980   &   700$\pm$40  \nl
1999 May   &  Speckle & 1960  &   740$\pm$110  \nl 
1999 May   &  AO       & 5     &  570$\pm$40   \nl
1999 July  & Speckle  & 1370  &   680$\pm$90  \nl
\hline
$<$Speckle$>$ &   &        &    725$\pm$40  \nl  
$<$Final$>$ &   &            &    650$\pm$80$^b$  \nl  
\enddata
\tablenotetext{a}{Radii are estimated for the individual AO images and SAA images each of which is composed of 100 frames. Here
we report the average and standard deviation of these estimates from each epoch.}
\tablenotetext{b}{This size is the average of the radii estimated from the SAA and AO data and the uncertainty is the half-range of these
values.}
\end{deluxetable}

\begin{deluxetable}{cccccccc} 
\tablecaption{Mid-infrared Radius and Flux of IRS 21} 
\tablewidth{43pc}
\tablehead{
\colhead{$\lambda$}   &\colhead{Standard}&\colhead{PSF Star}&\colhead{$\theta_{PSF}$}&\colhead{Radius}& \colhead{A$_\lambda^b$ } &\colhead{Total Flux$^c$}&\colhead{Flux-Background$^c$} \\
\colhead{[$\micron$]} &\colhead{}        &\colhead{}        &\colhead{['']}         &\colhead{[AU]}   &\colhead{}                &\colhead{[Jy]}          &\colhead{[Jy]} 
} 
\startdata
  8.8   & IRS 7 \& 3   & IRS 3        &  0.59 &$<$1830$^a$ & 2.5  &3.21$\pm$1.9 &1.34$\pm$0.8 \nl
  12.5  & $\alpha$ Boo & $\alpha$ Boo &  0.31 &1550$\pm$140& 1.9  &40.3$\pm$2.7 &6.19$\pm$0.42 \nl
  20.8  & $\alpha$ Boo & $\alpha$ Boo &  0.52 &1480$\pm$200& 2.0  &58.0$\pm$17 &3.70$\pm$1.10 \nl
  24.5  & $\alpha$ Sco & $\alpha$ Sco &  0.61 &1880$\pm$160& 1.5  &78.3$\pm$24 &4.93$\pm$1.50 \nl
\enddata
\tablenotetext{a}{This radius represents a 3 $\sigma$ upper limit.}
\tablenotetext{b}{The ISM extinction, A$_\lambda$, is derived from the extinction law of Moneti et al. 2001}
\tablenotetext{c}{Observed flux densities are within a 1$\farcs$0 radius circular aperture.}
\end{deluxetable}

\begin{deluxetable}{ccccccccc}
\tablecaption{Best Fitting Parameters} 
\tablewidth{40pc}
\tablehead{ 
\colhead{Model} & \colhead{$\tilde{\chi}^2$} & \colhead{$\rho_d$}                &\colhead{T$_{cen}$} &\colhead{R$_{cen}$} & \colhead{L$_{cen}$}           &\colhead{$\rho_{NA}$}              & \colhead{r$_i$} & \colhead{r$_o$} \\
\colhead{}      & \colhead{}                 &\colhead{10$^{-17}$ [g cm$^{-3}$]} &\colhead{[K]}       &\colhead{[AU]}      & \colhead{10$^{4}$ [L$_{\sun}$]} &\colhead{10$^{-18}$ [g cm$^{-3}$]} &\colhead{[AU]}   &\colhead{[AU]} 
}

\startdata
Constant & 800  & 1.01 & 1000 & 17 & 1.2   & ...  & 500 & ...  \nl 
1/r$^2$  &  80  & 0.12 & 980  & 20 & 1.5   & 0.95 & 780 & 1210 \nl
Bowshock &  26  & 2.3  & 970  & 19 & 1.3   & 7.1  & 895 & 1800 \nl 
\enddata
\end{deluxetable}

\begin{deluxetable}{ccccccc}
\tablecaption{Properties of Stars with Luminosities of 10$^4$ L$_{\sun}$} 
\tablewidth{40pc}
\tablehead{     
\multicolumn{4}{c}{Observed Properties}                                                                                     &\vline& \multicolumn{2}{c}{Modeled Properties$^a$}          \\
              \colhead{Type}  & \colhead{L}               & \colhead{$\dot{m_*}^b$}           & \colhead{v$_{w}^c$}         &\vline& \colhead{1/r$^2$ - $\dot{m}$}  & \colhead{bowshock - $l$}  \\
              \colhead{}      & \colhead{10$^4$ L$_{\sun}$} & \colhead{[M$_{\sun}$ yr$^{-1}$]}  & \colhead{[km/s]}            &\vline& \colhead{[M$_{\sun}$ yr$^{-1}$]} & \colhead{[AU]} 
}
\startdata
A5 I       & 3.5    & 1.96$\times10^{-7}$  & 180$^b$  & \vline &2.1$\times10^{-4}$ & 70\nl
G0 I       & 3.0    & 3.06$\times10^{-9}$  & 160      & \vline& 2.0$\times10^{-4}$ & 8 \nl
K0 I       & 2.9    & 4.22$\times10^{-7}$  & 20       & \vline& 2.3$\times10^{-5}$& 33 \nl
K5 Iab     & 0.7    & 3.0$\times$10$^{-8}$ & 60      & \vline & 7.2$\times10^{-5}$ & 16\nl 
M1.5 I$^d$ & 4.8    & 10$^{-6}$            & 17      & \vline & 2.0$\times10^{-5}$ & 50\nl 
AGB$^d$    & 0.6    & 10$^{-5}$          & 40       &\vline  & 3.5$\times10^{-5}$ & {\bf 735 }\nl
WR$^d$     & 3-100  & 10$^{-6}$-3$\times$10$^{-4}$& 700-2500&\vline & 8.1-30$\times10^{-4}$ & {\bf 310-10,100} \nl
Ofpe/WN9$^e$ & 0.4-2.0 & 2-5$\times$10$^{-5}$ & 400    &\vline & 4.7$\times10^{-4}$  & {\bf 1270} \nl
WR WC9$^d$ & 28     & 10$^{-5}$             & 1000 &\vline & 1.2$\times10^{-3}$ & {\bf 1160} \nl
\enddata
\tablenotetext{a}{Modeled values are to be compared with observed values for $\dot{m}$ (column 3) and size ($\sim$ 1000 AU). 
Those modeled properties which agree with the observations are in bold type. This rules out the mass-loss scenario and supports a bow shock model (column 6) for the 
last four entries.}
\tablenotetext{b}{Lang (1997) unless otherwise noted}
\tablenotetext{c}{Kudritzki \& Puls (2000) unless otherwise noted}
\tablenotetext{d}{Lamers \& Cassinelli (1999)}
\tablenotetext{e}{Pasquali et al. (1997)}
\end{deluxetable}
\end{document}